\documentclass[twocolumn,twoside,slac]{revtex4}
\usepackage{graphicx}
\usepackage{fancyhdr}
\begin{document}

\title{Constructing Ensembles of Pseudo-Experiments}
\author{Luc Demortier}
\affiliation{The Rockefeller University, New York, NY 10021, USA}

\begin{abstract}
The frequentist interpretation of measurement results requires the
specification of an ensemble of independent replications of the
same experiment. For complex calculations of bias, coverage,
significance, etc., this ensemble is often simulated by running
Monte Carlo pseudo-experiments. In order to be valid, the latter
must obey the Frequentist Principle and the Anticipation Criterion.
We formulate these two principles and describe some of their consequences
in relation to stopping rules, conditioning, and nuisance parameters.
The discussion is illustrated with examples taken from high-energy
physics.
\end{abstract}

\maketitle

\thispagestyle{fancy}

%%%%%%%%%%%%%%%%%%%%%%%%%%%%%%%%%%%%%%%%%%%%%%%%%%%%%%%%%%%%%%%%%%%%%%%%%%%%%%%%%
\section{Introduction}
Many statistical analyses in physics are based on a frequency interpretation 
of probability.  For example, the result of measuring a physical constant 
$\theta$ can be reported in the form of a $1-\alpha$ confidence interval 
$[X_{1},X_{2}]$, with the understanding that if the measurement is replicated 
a large number of times, one will have $X_{1}\le\theta\le X_{2}$
in a fraction $1-\alpha$ of the replications.  This type of interpretation 
therefore requires the definition of a {\em reference set} of 
similar measurements:
\begin{quote}{\em
The reference set of a measurement is the ensemble of experiments in
which the actually performed experiment is considered to be embedded
for the purpose of interpreting its results in a frequentist framework.
}\end{quote}
A major appeal of frequentism among physicists is its empirical definition 
of probability.  By the strong law of large numbers, probabilities can be 
approximated in finite ensembles, and such approximations converge to the 
true value as the ensemble size increases.  In other words, frequentist 
confidence statements are experimentally verifiable.

Physicists use Monte Carlo generated ensembles in various applications:
to check a fitting algorithm for the presence of bias, non-Gaussian
pulls, or other pathologies; to calculate the coverage of confidence 
intervals or upper limits;  to average out statistical fluctuations in 
order to isolate systematic effects; to calculate goodness-of-fit 
measures and sig\-nifi\-cances; to design experiments; etc.
When constructing ensembles to address these questions, one needs to
pay attention to a number of subtle issues that arise in a frequentist
framework: what is the correct stopping rule?; is it appropriate to
condition, and if so, on what statistic?; how should nuisance parameters
be handled?

The aim of this paper is to draw attention to these issues and to propose
some recommendations where possible.  We start by discussing basic frequentist
principles in section \ref{FreqPrin} and illustrate them with an example of
conditioning in section \ref{Cond}.  The importance of stopping rules is argued
in section \ref{StopRule}.  Finally, some purely frequentist methods to handle 
nuisance parameters are described in section \ref{NuisPara}.

%%%%%%%%%%%%%%%%%%%%%%%%%%%%%%%%%%%%%%%%%%%%%%%%%%%%%%%%%%%%%%%%%%%%%%%%%%%%%%%%%
\section{Frequentist Principles}
\label{FreqPrin}
%%%%%%%%%%%%%%%%%%%%%%%%%%%%%%%%%%%%%%%%%%%%%%%%%%%%%%%%%%%%%%%%%%%%%%%%%%%%%%%%%
In order to deserve the label frequentist, a statistical procedure and its
associated ensemble must satisfy two core principles, which we examine in
the next two subsections.

%%%%%%%%%%%%%%%%%%%%%%%%%%%%%%%%%%%%%%%%%%%%%%%%%%%%%%%%%%%%%%%%%%%%%%%%%%%%%%%%%
\subsection{The Frequentist Guarantee}
\label{FreqGuar}
%%%%%%%%%%%%%%%%%%%%%%%%%%%%%%%%%%%%%%%%%%%%%%%%%%%%%%%%%%%%%%%%%%%%%%%%%%%%%%%%%
The first principle states the aims of frequentism:
\begin{quote}{\bf Frequentist Guarantee \protect\cite{baya2003}:}\\
{\em In repeated use of a statistical procedure, the long-run average 
actual accuracy should not be less than (and ideally should equal) 
the long-run average reported accuracy.
}\end{quote}
To clarify this principle, we return to the $1-\alpha$ confidence interval
procedure mentioned in the Introduction.  Let ${\cal E}$ be an ensemble
of intervals obtained by applying this procedure many times on different, 
independent data.  The {\em actual accuracy} of an interval in ${\cal E}$ 
is $1$ or $0$: either the interval covers the true value of the parameter 
of interest, or it does not.  The average actual accuracy is therefore simply 
the fraction of intervals in ${\cal E}$ that cover.
On the other hand, the average {\em reported accuracy} is $1-\alpha$.
The reported accuracy is often the same for all intervals in ${\cal E}$,
but in some settings it is possible to report a different, data-dependent 
accuracy for each interval.  Thus, averaging the reported accuracy is not 
necessarily a trivial operation.  A procedure that satisfies the Frequentist 
Guarantee is said to have coverage.

In a sense, the Frequentist Guarantee is only weakly constraining, 
because it does not require a procedure to have coverage when applied to 
repeated measurements of the {\em same} quantity.  To see how this is
relevant, consider the construction of a 68\% confidence interval for the 
mean $\mu$ of a Poisson distribution.  One procedure is to take all $\mu$ 
values satisfying $(n-\mu)^{2}/\mu\leq 1$, where $n$ is the observed 
number of events.  The resulting interval actually undercovers for many
values of $\mu$ and overcovers for other values, so that the Frequentist
Guarantee appears to be satisfied {\em on average}.  To make this 
statement more precise we need a weighting function with which to carry
out the average over $\mu$.  A simple proposal is to perform local smoothing
of the coverage function, resulting in local average coverage \cite{baya2003}.  

Physicists may object to this notion of local average coverage on the
grounds that they sometimes repeatedly measure a given constant of nature
and are then interested in the coverage obtained for that particular 
constant, not in an average coverage over ``nearby'' constants.  A possible 
answer is that one rarely measures the quantity of interest directly.  
Rather, one measures a combination of the quantity of interest with 
calibration constants, efficiencies, sample sizes, etc., all of which vary 
from one measurement to the next, so that an effective averaging does take 
place.

Finally, it could be argued that even Bayesians should subscribe to some 
form of the Frequentist Guarantee.  If, over repeated use, a 95\% credible 
Bayesian interval fails to cover the true value more than 30\% of the time 
(say), then there must be something seriously wrong with that interval.

%%%%%%%%%%%%%%%%%%%%%%%%%%%%%%%%%%%%%%%%%%%%%%%%%%%%%%%%%%%%%%%%%%%%%%%%%%%%%%%%%
\subsection{The Anticipation Criterion}
\label{AntiCrit}
%%%%%%%%%%%%%%%%%%%%%%%%%%%%%%%%%%%%%%%%%%%%%%%%%%%%%%%%%%%%%%%%%%%%%%%%%%%%%%%%%
Although the Frequentist Guarantee specifies how a statistical procedure 
should behave under many repetitions of a measurement, it does not indicate 
what constitutes a valid repetition, and hence a valid ensemble.  To the extent
that this question involves the notion of randomness, it is well beyond the 
scope of this paper.  From a practical standpoint however, one would like 
to stipulate that all effects susceptible to interfere with that randomness 
must be recognized as such and included in the construction of the ensemble, 
i.e. ``anticipated''\cite{brow1977}.  Hence the second principle:
\begin{quote}{\bf Anticipation Criterion:}\\
{\em Ensembles must anticipate all elements of chance and all
elements of choice of the actual experiments they serve to interpret.
}\end{quote}
To clarify, ``elements of chance'' refers to statistical fluctuations of 
course, but also to systematic uncertainties when the latter come from nuisance 
parameters that are determined by auxiliary measurements.  On the other hand,
``elements of choice'' refers to actions by experimenters, in particular how 
they decide to stop the experiment, and what decisions they make after 
stopping.

One can identify several levels of anticipation.  At the highest level, the 
data collection and analysis methods, as well as the reference ensemble used 
to interpret results, are fully specified at the outset.  They do not change 
once the data is observed.  The reference ensemble is called ``unconditional''.

At the second highest level, the data collection and analysis methods are 
fully specified at the outset, but the reference ensemble is not.   The latter
will be fully determined once the data is observed, and is therefore
``conditional''.  Although a conditional ensemble is not known before observing 
the data, it is a subset in a known partition of a known unconditional ensemble.

The lowest level of anticipation is occupied by Bayesian methods, 
which fully condition on the observed data.  The reference ensemble collapses
to a point and can therefore no longer be used as a reference.

As the level of anticipation decreases, the reference ensemble becomes smaller.  
A remarkable result is that within the second level of
anticipation one can refine the conditioning partition to the point where it 
is possible to give a Bayesian interpretation to frequentist conclusions, and 
vice-versa \cite{jber1997}.

%%%%%%%%%%%%%%%%%%%%%%%%%%%%%%%%%%%%%%%%%%%%%%%%%%%%%%%%%%%%%%%%%%%%%%%%%%%%%%%%%
\section{Conditioning}
\label{Cond}
%%%%%%%%%%%%%%%%%%%%%%%%%%%%%%%%%%%%%%%%%%%%%%%%%%%%%%%%%%%%%%%%%%%%%%%%%%%%%%%%%
To illustrate the interplay between anticipation and conditioning, we present 
here a famous example originally due to Cox \cite{cox1958}.  Suppose we make 
one observation of a rare particle and wish to estimate its mass $\mu$ from 
the momenta of its decay products.  For the sake of simplicity, assume that 
the estimator $X$ of $\mu$ is normal with mean $\mu$ and variance $\sigma^{2}$.  
There is a 50\% chance that the particle decays hadronically, in which case 
$\sigma=10$; otherwise the particle decays leptonically and $\sigma=1$.  
Consider the following 68\% confidence interval procedures:
\begin{enumerate}
\item \underline{Unconditional}\\
      If the particle decayed hadronically, report $X\pm \delta_{h}$,
      otherwise report $X\pm\delta_{\ell}$, where $\delta_{h}$ and
      $\delta_{\ell}$ are chosen so as to minimize the expected length
      $\langle\delta\rangle=\delta_{h}+\delta_{\ell}$ subject to the 
      constraint of 68\% coverage.  This yields $\delta_{\ell}=2.20$ 
      and $\delta_{h}=5.06$.  The expected length is 7.26.
\item \underline{Conditional}\\
      If we condition on the decay mode, then the best interval is
      $X\pm 10$ if the particle decayed hadronically, and $X\pm 1$
      otherwise.  So the expected length is 11.0 in this case.
\end{enumerate}
Note that in both cases we used all the information available: the 
measurement $X$ as well as the decay mode.  Both procedures are valid;  
the only difference between them is the reference frame.  
The unconditional ensemble includes both decay modes, whereas the 
conditional one only includes the observed decay mode.

The expected length is shorter for unconditional intervals than for 
conditional ones.  Does this mean we should quote the former?  If our 
aim is to report what we learned from the data we observed, then clearly
we should report the conditional interval.  Suppose indeed that we observed
a hadronic decay.  The unconditional interval width is then 10.12, compared
to 20.0 for the conditional one.  The reason the unconditional interval is 
shorter is that, if we could repeat the experiment, we might observe the 
particle decaying into the leptonic mode.  However, this is irrelevant
to the interpretation of the observation we actually made.  
This example illustrates a general feature of conditioning, that it usually 
increases expected length, and reduces power in test settings.

Another aspect of the previous example is that the conditioning statistic
(the decay mode) is ancillary: its distribution does not depend on the 
parameter of interest (the particle mass).  This is not always the case.
Suppose for example that we are given a sample from a normal distribution
with unit variance and unknown mean $\theta$, and that we wish to test
$H_{0}: \theta=-1$ versus $H_{1}: \theta=+1$.  The standard symmetric Neyman-Pearson
test based on the sample mean $\bar{X}$ as test statistic rejects $H_{0}$
if $\bar{X}>0$.  It makes no distinction between $\bar{X}=0.5$ and $\bar{X}=5$,
even though in the latter case we certainly feel more confident in our 
rejection of $H_{0}$.  Although $\bar{X}$ is not ancillary, it is possible
to use it to calculate a conditional ``measure of confidence'' to help 
characterize one's decision regarding $H_{0}$ \cite{kief1977}.  Unfortunately, 
a general theory for choosing such conditioning statistics does not exist.

%%%%%%%%%%%%%%%%%%%%%%%%%%%%%%%%%%%%%%%%%%%%%%%%%%%%%%%%%%%%%%%%%%%%%%%%%%%%%%%%%
\section{Stopping Rules}
\label{StopRule}
%%%%%%%%%%%%%%%%%%%%%%%%%%%%%%%%%%%%%%%%%%%%%%%%%%%%%%%%%%%%%%%%%%%%%%%%%%%%%%%%%
Stopping rules specify how an experiment is to be terminated.
High-energy physics experiments are often sequential, so it is
important to properly incorporate stopping rules in the construction
of ensembles.  

As a first example, consider the measurement of the branching fraction 
$\theta$ for the decay of a rare particle $A$ into a particle $B$.  
Suppose we observe a total of $n=12$ decays, $x=9$ of which are 
$A\rightarrow B$ transitions, and the rest, $r=3$, are $A\not\rightarrow B$
transitions.  We wish to test $H_{0}: \theta=1/2$ versus $H_{1}: \theta>1/2$.

A possible stopping rule is to stop the experiment after observing a total 
number of decays $n$.  The probability mass function (pmf) is then binomial:
\begin{equation}
f(x\,;\,\theta)\;=\;\left(\begin{array}{c}
                          n \\ x
                          \end{array}\right) 
                    \theta^{x}\;(1-\theta)^{n-x},
\end{equation}
and the $p$ value for testing $H_{0}$ is:
\begin{equation}
p_{b} \;=\; \sum_{i=9}^{12}\left(\begin{array}{c} 12 \\ i 
\end{array}\right)\theta^{i}\;(1-\theta)^{12-i}\;=\;0.075.
\end{equation}

An equally valid stopping rule is to stop the experiment after observing a 
number $r$ of $A\not\rightarrow B$ decays.  Now the pmf is negative binomial:
\begin{equation}
f(x\,;\,\theta) \;=\;\left(\begin{array}{c}
                           r+x-1 \\ x
                           \end{array}\right)
                     \theta^{x}\;(1-\theta)^{r},
\end{equation}
and the $p$ value is:
\begin{equation}
p_{nb} \;=\; \sum_{i=9}^{\infty}\left(\begin{array}{c} 2+i \\ i 
\end{array}\right)\theta^{i}\;(1-\theta)^{3}\;=\;0.0325.
\end{equation}

If we adopt a 5\% threshold for accepting or rejecting $H_{0}$, 
we see that the binomial model leads to acceptance,
whereas the negative binomial model leads to rejection.

Here is a more intriguing example \cite{jber1988}.  Imagine a physicist 
working at some famous particle accelerator and developping a procedure 
to select collision events that contain a Higgs boson.  Assume that the 
expected rate of background events accepted by this procedure is known 
very accurately.  Applying his technique to a given dataset,
the physicist observes $68$ events and expects a background of $50$.  The
(Poisson) probability for 50 to fluctuate up to 68 or more is 0.89\%,
and the physicist concludes that there is significant evidence against
$H_{0}$, the background-only hypothesis, at the 1\% level.

Is this conclusion correct?  Perhaps the physicist just decided to take 
a single sample.  But what would he have done if this sample had not 
yielded a significant result?  Perhaps he would have taken another sample!  
So the real procedure the physicist was considering is actually of the form:
\begin{itemize}
\item Take a data sample, count the number $n_{1}$ of Higgs candidates,
      and calculate the expected background $b$;
\item If I$\!$P$(N\ge n_{1}\,|\,b)\le\alpha$ then stop and reject $H_{0}$;
\item Otherwise, take a second sample with the same expected background, 
      count the number $n_{2}$ of Higgs candidates and reject $H_{0}$ if
      I$\!$P$(N\ge n_{1}+n_{2}\,|\,2b)\le\alpha$.
\end{itemize}
For this test procedure to have a level of 1\%, $\alpha$ must be set at
0.67\%.  Since the {\em actual} data had a $p$ value of 0.89\%, the physicist 
should not have rejected $H_{0}$.

So now the physicist finds himself forced to take another sample.
There are two interesting cases:
\begin{enumerate}
\item The second sample yields 57 candidate events, for a total of 125.
      The probability for the expected background (100 events now) to 
      fluctuate up to 125 or more is $0.88\% > 0.67\%$, so the result is 
      not significant.  However, the result would have been significant
      if the physicist had not stopped halfway through data taking to
      calculate the $p$ value!
\item The second sample yields 59 candidate events, for a total of 127.
      The $p$ value is now 0.52\% and significance has been obtained,
      unless of course the physicist was planning to take a third sample 
      in the event of no significance.
\end{enumerate}
Bayesian methods are generally independent of the stopping rule.  It is
therefore somewhat ironic that frequentists, who start from an objective 
definition of probability, should end up with results that depend on the 
thought processes of the experimenter.

%%%%%%%%%%%%%%%%%%%%%%%%%%%%%%%%%%%%%%%%%%%%%%%%%%%%%%%%%%%%%%%%%%%%%%%%%%%%%%%%%
\section{Nuisance Parameters}
\label{NuisPara}
%%%%%%%%%%%%%%%%%%%%%%%%%%%%%%%%%%%%%%%%%%%%%%%%%%%%%%%%%%%%%%%%%%%%%%%%%%%%%%%%%
Most problems of inference involve nuisance parameters, i.e. uninteresting
parameters that are incompletely known and therefore add to the overall
uncertainty on the parameters of interest.
To fix ideas, assume that we have a sample $\{x_{1},\ldots,x_{n}\}$ whose 
probability density function (pdf) $f(\vec{x}\,;\,\mu,\nu)$ depends on a 
parameter of interest $\mu$ and a nuisance parameter $\nu$,
and that the latter can be determined from a separate sample
$\{y_{1},\ldots,y_{m}\}$ with pdf $g(\vec{y}\,;\,\nu)$.
Correct inference about $\mu$ must then be derived from the joint pdf
\begin{equation}
h(\vec{x},\vec{y}\,;\,\mu,\nu)\;\equiv\; f(\vec{x}\,;\,\mu,\nu)\;g(\vec{y}\,;\,\nu).
\end{equation}
What is often done in practive however, is to first obtain a distribution
$\pi(\nu)$ for $\nu$, usually by combining measurement results with a
sensible guess for the form of $\pi(\nu)$.  Inference about $\mu$ 
is then based on:
\begin{equation}
h^{\prime}(\vec{x}\,;\,\mu)\;\equiv\;\int f(\vec{x}\,;\,\mu,\nu)\,\pi(\nu)\,d\nu.
\label{eq:hybrid}
\end{equation}
Although this technique borrows elements from both Bayesian and frequentist
methodologies, it really belongs to neither and is more properly referred to
as a hybrid non-frequentist/non-Bayesian approach.

We illustrate the handling of nuisance parameters with a simple $p$ value 
calculation.  Suppose that a search for a new particle ends with a sample of 
$n_{0}=12$ candidates over a separately measured background of $\nu_{0}=5.7\pm 0.47$,
where we ignore the uncertainty on the standard error $0.47$. 
Let $\mu$ be the unknown expected number of new particles among
the 12 candidates.  We wish to test $H_{0}: \mu=0$ versus $H_{1}: \mu>0$.

A typical model for this problem consists of a Poisson density for the number
of observed candidates and a Gaussian for the background measurement.
Using equation (\ref{eq:hybrid}) with a simple Monte Carlo integration
routine, one obtains a $p$ value of $\sim 1.6\%$.
For reference, when there is no uncertainty on $\nu_{0}$ the $p$
value is $\sim1.4\%$.

While there are many purely frequentist approaches to the elimination
of nuisance parameters, few of these have general applicability.  
Concentrating on the latter, we discuss the likelihood ratio and confidence 
interval methods in the next two subsections.

%%%%%%%%%%%%%%%%%%%%%%%%%%%%%%%%%%%%%%%%%%%%%%%%%%%%%%%%%%%%%%%%%%%%%%%%%%%%%%%%%
\subsection{Likelihood Ratio Method}
The likelihood ratio statistic $\lambda$ is defined by:
\begin{equation}
\lambda \;=\;
\frac{\begin{array}[b]{l}
      \sup\; {\cal L}(\mu,\nu\,|\,n_{0},\nu_{0})\\[-4pt]
      {\scriptstyle \mu  = 0}       \\[-4pt]
      {\scriptstyle \nu\ge 0}
      \end{array}}
     {\begin{array}[t]{l}
      \sup\; {\cal L}(\mu,\nu\,|\,n_{0},\nu_{0})\\[-4pt]
      {\scriptstyle \mu\ge 0}       \\[-4pt]
      {\scriptstyle \nu\ge 0}
     \end{array}}
,
\label{eq:LikelihoodRatio}
\end{equation}
where, for $\nu_{0}\gg\Delta\nu$:
\begin{displaymath}
{\cal L}(\mu,\nu\,|\,n_{0},\nu_{0})\;\propto\; 
\frac{(\mu+\nu)^{n_{0}}}{n_{0}!}\,\,
e^{-\mu-\nu}\;
e^{-\frac{1}{2}\left(\frac{\scriptstyle \nu-\nu_{0}}{\scriptstyle\Delta\nu}\right)^{2}}.
\end{displaymath}
Simple calculus leads to:
\begin{displaymath}
\begin{array}{cclc}
-2\ln\lambda & = & 2\left(n_{0}\ln\frac{n_{0}}{\hat{\nu}}+\hat{\nu}-n_{0}\right)+
                   \left(\frac{\hat{\nu}-\nu_{0}}{\Delta\nu}\right)^{2} 
                 & \textrm{if}\; n_{0}>\nu_{0},\\[1.5mm]
             & = & 0 
                 & \textrm{if}\; n_{0}\le\nu_{0},\\[2mm]
\multicolumn{4}{l}{\textrm{with:}\; \hat{\nu}\;=\;\frac{\nu_{0}-\Delta\nu^{2}}{2}+
\sqrt{\left(\frac{\nu_{0}-\Delta\nu^{2}}{2}\right)^{2}+n_{0}\,\Delta\nu^{2}}.}
\end{array}
\end{displaymath}
Since $\lambda$ depends on $n_{0}$ and $\nu_{0}$, its distribution under $H_{0}$
depends on the true expected background $\nu_{t}$.  A natural simplification is 
to examine the limit $\nu_{t}\rightarrow\infty$.  Application of theorems 
describing the asymptotic behavior of $-2\ln\lambda$ must take into account 
that for $n_{0}<\nu_{0}$ the analytical maximum of the likelihood lies outside 
the physical region $\mu\ge 0$.  The correct asymptotic result is that, under 
$H_{0}$, half a unit of probability is carried by the singleton 
$\{-2\ln\lambda=0\}$, and the other half is distributed as a chisquared with one
degree of freedom over $0<-2\ln\lambda<+\infty$.

For our example the expected background is only 5.7 particles however, 
so one may wonder how close this is to the asymptotic limit.  Here is an 
algorithm to check this.  Choose a true number of background events $\nu_{t}$ 
and repeat the following three steps a large number of times:
\begin{enumerate}
\addtolength{\itemsep}{-2mm}
\item Generate a Gaussian variate $\nu_{0}$ with mean $\nu_{t}$ and width $\Delta\nu$;
\item Generate a Poisson variate $n_{0}$ with mean $\nu_{t}$;
\item Calculate $\lambda$ from the generated $\nu_{0}$ and $n_{0}$.
\end{enumerate}
The $p$ value is then equal to the fraction of pseudo-experiments that 
yield a likelihood ratio $\lambda$ smaller than the $\lambda_{0}$ obtained 
from the observed data.

Note that this algorithm does not ``smear'' the true value of any parameter,
in contrast with equation (\ref{eq:hybrid}).  The price for this is
that the result depends on the choice of $\nu_{t}$.   For $\nu_{t}$ varying
from $0.5$ to $50$, the $p$ value ranges from $\sim 0.48$ to $\sim 1.2\%$.  
A general prescription for dealing with a $p$ value dependence on nuisance
parameters is to use the so-called supremum $p$ value:
\begin{displaymath}
p_{\sup}\;=\;\sup_{\nu}\; \left.\mbox{\rm I$\!$P}
(-2\ln\lambda \ge -2\ln\lambda_{0}\,|\,\mu,\nu)\right|_{\mu=0}
\end{displaymath}
From a frequentist point of view, the supremum $p$ value is {\em valid}, in the
sense that:
\begin{equation}
\mbox{\rm I$\!$P}(p_{\sup}\le\alpha)\;\le\;\alpha, 
\;\;\;\mbox{\rm for each}\;\alpha\in [0,1],
\label{eq:valid}
\end{equation}
regardless of the true value of the nuisance parameter.  Although it is often
difficult to calculate a supremum, in this case it turns out to equal the
asymptotic limit to a good approximation.  In our example $-2\ln\lambda_{0}=5.02$ 
and corresponds to $p_{\sup} \approx p_{\infty} = 1.25\%$.

As the attentive reader will have noticed, the $p$ value is smaller for 
$\Delta\nu=0.47$ than for $\Delta\nu=0$.  This is a consequence of the 
discreteness of Poisson statistics; it does not violate inequality (\ref{eq:valid})
because $p_{\sup}$ actually overcovers a little when $\Delta\nu=0$.  To avoid the 
bias resulting from this overcoverage, the use of mid-$p$ values is sometimes 
advocated for the purpose of comparing or combining $p$ values \cite{berr1995}.

%%%%%%%%%%%%%%%%%%%%%%%%%%%%%%%%%%%%%%%%%%%%%%%%%%%%%%%%%%%%%%%%%%%%%%%%%%%%%%%%%
\subsection{Confidence Interval Method}
The supremum $p$ value introduced in the previous section can be defined
for any test statistic, although it will not always give useful results.
If for example in our new particle search we take the total number $n_{0}$ of 
observed candidates as test statistic, the $p$ value will be 100\% since the
background $\nu$ is unbounded from above.  A more satisfactory
method proceeds as follows \cite{berg1994,silv1996}.  First, construct a 
$1-\beta$ confidence interval $C_{\beta}$ for the nuisance parameter $\nu$, 
then maximize the $p$ value over that interval, and finally correct the result 
for the fact that $\beta\ne 0$:
\begin{displaymath}
p_{\beta}\;=\;\sup_{\nu\in C_{\beta}}\; 
\left.\mbox{\rm I$\!$P}(N\ge n_{0}\,|\,\mu,\nu)\right|_{\mu=0}\;+\;\beta.
\end{displaymath}
It can be shown that this is also a {\em valid} $p$ value.

For the sake of illustration with our example, we consider three choices
of $\beta$ and construct the corresponding $1-\beta$ confidence intervals
for $\nu_{t}$:
\begin{displaymath}
\begin{array}{llll}
1-\beta= & 99.5\%:  & C_{0.005}  & =\;[4.38\,,\,7.02] \\
1-\beta= & 99.9\%:  & C_{0.001}  & =\;[4.15\,,\,7.25] \\
1-\beta= & 99.99\%: & C_{0.0001} & =\;[3.87\,,\,7.53]
\end{array}
\end{displaymath}
To calculate the $p$ value, a good choice of statistic is the
maximum likelihood estimator of the signal, i.e. $\hat{s}\equiv n_{0}-\nu_{0}$.
Under $H_{0}$, the survivor function of $\hat{s}$ is given by:
\begin{displaymath}
\mbox{\rm I$\!$P}(S\ge \hat{s})\;=\;
\sum_{k=0}^{\infty}
\frac{1+\textrm{erf}\left(\frac{k-\nu_{t}-\hat{s}}{\sqrt{2}\,\Delta\nu}\right)}
     {1+\textrm{erf}\left(\frac{\nu_{t}}{\sqrt{2}\,\Delta\nu}\right)}\:
\frac{\nu_{t}^{k}}{k!}\:e^{-\nu_{t}}
\end{displaymath}
We then find:
\begin{displaymath}
\begin{array}{lll@{}ll}
1-\beta= &  99.5\%:  & p_{\beta}\;=\;1.6\%  & +0.5\%  &  =\; 2.1\% \\
1-\beta= &  99.9\%:  & p_{\beta}\;=\;1.7\%  & +0.1\%  &  =\; 1.8\% \\
1-\beta= &  99.99\%: & p_{\beta}\;=\;1.88\% & +0.01\% &  =\; 1.89\%
\end{array}
\end{displaymath}

An important point about the confidence interval method is that, in order to 
satisfy the Anticipation Criterion, the value of $\beta$ and the confidence 
set $C_{\beta}$ must be specified before looking at the data.  Since $p_{\beta}$ 
is never smaller than $\beta$, the latter should be small.
In particular, if $p_{\beta}$ is used in a level-$\alpha$ test, then $\beta$
must be smaller than $\alpha$ for the test to be useful.  

%%%%%%%%%%%%%%%%%%%%%%%%%%%%%%%%%%%%%%%%%%%%%%%%%%%%%%%%%%%%%%%%%%%%%%%%%%%%%%%%%
\section{Summary}
%%%%%%%%%%%%%%%%%%%%%%%%%%%%%%%%%%%%%%%%%%%%%%%%%%%%%%%%%%%%%%%%%%%%%%%%%%%%%%%%%
From the practical point of view of someone analyzing data, the most critical 
property of frequentist ensembles is their ``anticipatoriness.''  This requires
that all the structural elements of an analysis (i.e. test sizes, interval
procedures, bin boundaries, stopping rules, etc.) be in place before looking at
the data.  The only exception to this requirement occurs in situations where
conditioning is both possible and appropriate.  Even in that case, the conditioning
partition itself must be specified beforehand.

%%%%%%%%%%%%%%%%%%%%%%%%%%%%%%%%%%%%%%%%%%%%%%%%%%%%%%%%%%%%%%%%%%%%%%%%%%%%%%%%%

%%%%%%%%%%%%%%%%%%%%%%%%%%%%%%%%%%%%%%%%%%%%%%%%%%%%%%%%%%%%%%%%%%%%%%%%%%%%%%%%%
\end{document}